\documentclass[a4paper]{jpconf}  
\usepackage[dvips]{graphicx}
\usepackage{dcolumn}   
\usepackage{bm}        
\usepackage{amssymb}   
\usepackage[utf8]{inputenc}
\usepackage{mathrsfs}
\usepackage{amsmath}
\usepackage[english]{babel}
\usepackage{hyperref}
\usepackage{textcomp}
\usepackage{subfigure}
\usepackage{color}
\usepackage{cite}
\begin{document}

\title{A new look at the double sine-Gordon kink-antikink scattering}

\author{E~Belendryasova$^{1}$, V~A~Gani$^{1,2}$, A~Moradi Marjaneh$^3$, D~Saadatmand$^4$ and A~Askari$^5$}

\address{$^1$National Research Nuclear University MEPhI (Moscow Engineering Physics Institute), 115409 Moscow, Russia}

\address{$^2$National Research Center Kurchatov Institute, Institute for Theoretical and Experimental Physics, 117218 Moscow, Russia}

\address{$^3$Young Researchers and Elite Club, Quchan Branch, Islamic Azad university, Quchan, Iran}

\address{$^4$Department of Physics, University of Sistan and Baluchestan, Zahedan, Iran}

\address{$^5$Department of Physics, Faculty of Science, University of Hormozgan, P.O.Box 3995, Bandar Abbas, Iran}

\ead{vagani@mephi.ru}

\begin{abstract}

We study the kink-antikink scattering within the double sine-Gordon model. In the numerical simulations we found a critical value $v_\mathrm{cr}$ of the initial velocity $v_\mathrm{in}$, which separates two different scenarios: at $v_\mathrm{in}<v_\mathrm{cr}$ the kinks capture each other and form a bound state, while at $v_\mathrm{in}>v_\mathrm{cr}$ the kinks pass through each other and escape to infinities. We obtain non-monotonous dependence of $v_\mathrm{cr}$ on the model parameter $R$. Besides that, at some initial velocities below $v_\mathrm{cr}$ we observe formation and interaction of the so-called oscillons (new phenomenon), as well as escape windows (well-known phenomenon).

\end{abstract}

\section{Introduction}

Field-theoretic models with polynomial and non-polynomial potentials are of great importance for various physical systems, e.g., in high energy physics, cosmology, condensed matter, etc., \cite{Rajaraman.book.1982,Vilenkin.book.2000,Manton.book.2004,Vachaspati.book.2006}. Significant progress has been made in studying topological solitons in models with one real scalar field \cite{lohe,Gani.PRE.1999,Gani.YaF.1999.eng,Gani.YaF.1999.rus,Gani.YaF.2001.eng,Gani.YaF.2001.rus,GaKuLi,Dorey.PRL.2011,khare,Gani.JHEP.2015,GaLeLiconf,Bazeia.EPJC.2017.conf.sinh,Bazeia.EPJC.2018.sinh,Belendryasova.conf.2017,Belendryasova.CNSNS.2019}, as well as in more complex models with two or more fields \cite{GaKu.SuSy.2001.eng,GaKu.SuSy.2001.rus,Lensky.JETP.2001.eng,Lensky.JETP.2001.rus,Kurochkin.CMMP.2004.eng,Kurochkin.CMMP.2004.rus,GaKsKu01.eng,GaKsKu01.rus,GaKsKu02.eng,GaKsKu02.rus,GaLiRa,GaLiRaconf,Radomskiy.JPCS.2017,GaKiRu,GaKiRu.conf}. Both analytical and numerical methods are successfully applied to studying kink-antikink interactions. In particular, the collective coordinate method with one or more degrees of freedom enables to model kink-(anti)kink interactions \cite{GaKuLi,Weigel.cc.2014,Demirkaya.cc.2017}. Many interesting results has been obtained by numerical modeling of collision of kink and antikink, of kink with a defect \cite{Saadat.CNSNS.2018}, and of several kinks in one point \cite{Moradi.JHEP.2017,Moradi.EPJB.2017,Saadatmand.PRD.2015}.

In this work we study the $(1+1)$-dimensional double sine-Gordon (DSG) model~\cite{Campbell.dsG.1986,Malomed.PLA.1989,Gani.PRE.1999,Gani.dsg.EPJC.2018}. Although this model has been well-investigated, we have obtained new results. Firstly, we have found a non-monotonous dependence of the critical velocity $v_\mathrm{cr}$ on the model parameter $R$. Secondly, at some initial velocities from the range $v_\mathrm{in}<v_\mathrm{cr}$ in the final state we observed complex oscillating structures --- oscillons. The latter phenomenon is rather new \cite{Bazeia.EPJC.2018.sinh,Bazeia.EPJC.2017.conf.sinh} and was not reported for the DSG model previously.

\section{Scattering of kinks}

\begin{figure}[h!]
\subfigure[potential]{
\begin{minipage}{0.45\linewidth}
\centering\includegraphics[width=\linewidth]{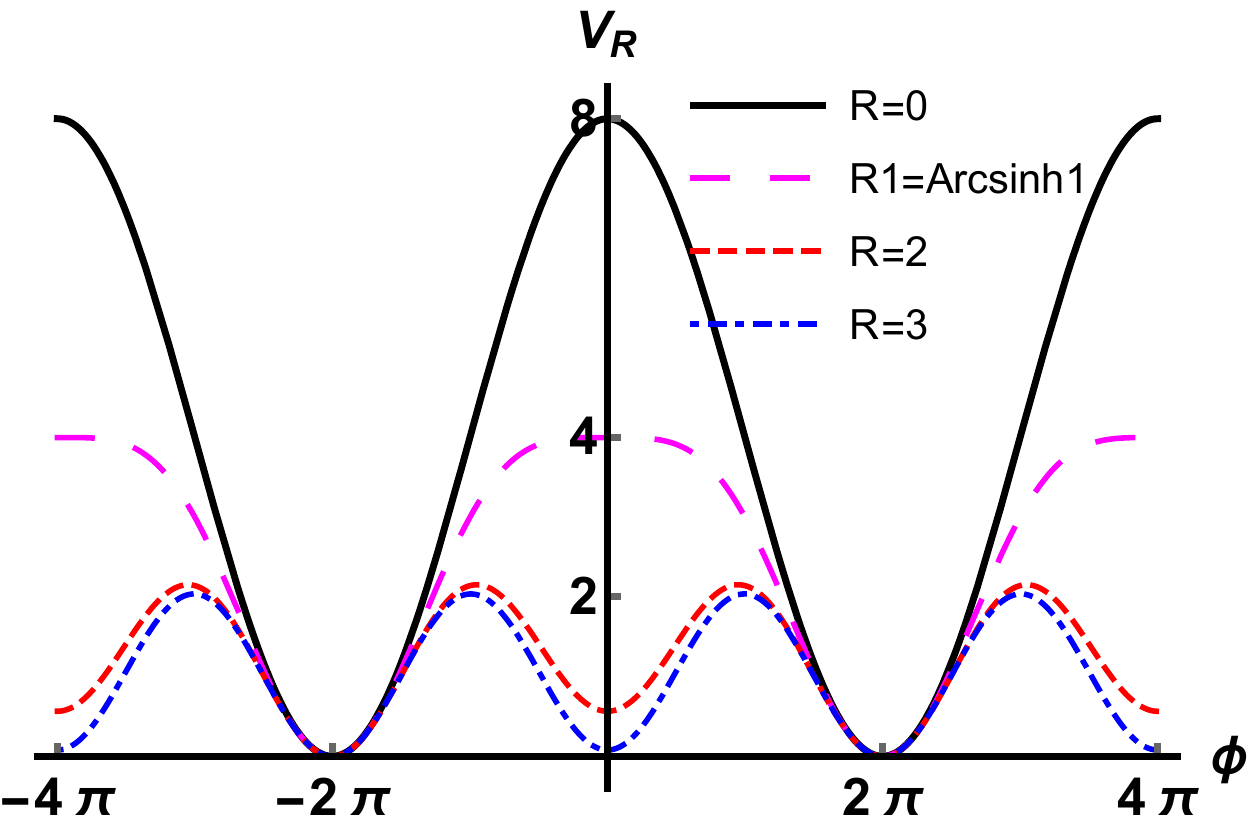}
\end{minipage}
}
\hfill
\subfigure[kinks]{
\begin{minipage}{0.45\linewidth}
\centering\includegraphics[width=\linewidth]{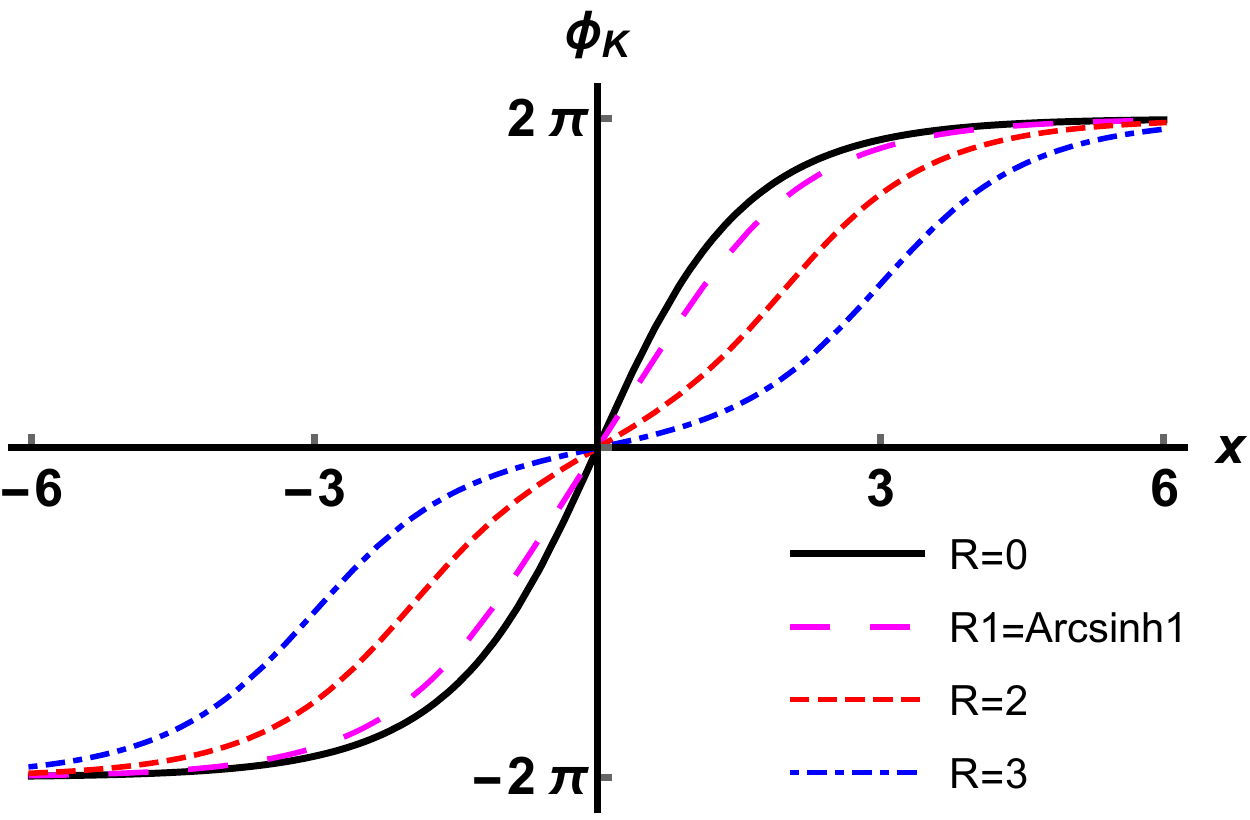}
\end{minipage}
}
\caption{Potential \eqref{eq:potential_R} and kinks \eqref{eq:DSG_kinks_1} from the sector $(-2\pi;2\pi)$ for some values of $R$.}
\label{fig:potential}
\end{figure}

Consider the $(1+1)$-dimensional double sine-Gordon model. Within this model the dynamics of a real scalar field $\phi(x,t)$ is described by the Lagrangian
\begin{equation}\label{eq:lagrangian}
\mathcal{L} = \frac{1}{2}\left(\frac{\partial \phi}{\partial t}\right)^2-\frac{1}{2}\left(\frac{\partial \phi}{\partial x}\right)^2-V(\phi)
\end{equation}
with the potential
\begin{equation}\label{eq:potential_R}
V_\mathrm{R}^{}(\phi) = \tanh^2R\:(1-\cos\phi) + \frac{4}{\cosh^2R}\left(1+\cos\frac{\phi}{2}\right), \qquad R>0,
\end{equation}
see figure \ref{fig:potential}a.
From the Lagrangian \eqref{eq:lagrangian} it is easy to obtain the equation of motion --- partial differential equation of the second order:
\begin{equation}\label{eq:eqmo}
\frac{\partial^2\phi}{\partial t^2}-\frac{\partial^2\phi}{\partial x^2}+\frac{dV}{d\phi}=0.
\end{equation}
In the static case $\phi=\phi(x)$, and we have the ordinary differential equation
\begin{equation}
\frac{d^2\phi}{dx^2}=\frac{dV}{d\phi}\quad\Leftrightarrow\quad \frac{d\phi}{dx}=\sqrt{2V},
\end{equation}
which yields static {\it kink} and {\it antikink} (see figure \ref{fig:potential}b):
\begin{equation}\label{eq:DSG_kinks_1}
\phi_{\mathrm{k}(\mathrm{\bar k})}(x) = 4\pi n \pm 4\arctan\frac{\sinh x}{\cosh R}.
\end{equation}
Note that the kink can be viewed as a superposition of two {\it subkinks} separated by distance $R$.

We performed numerical simulations of the kink-antikink scattering. To do this we solved the equation of motion \eqref{eq:eqmo} numerically using finite-difference method with the initial condition in the form of kink and antikink which were initially placed at $x=-\xi$ and $x=+\xi$, respectively, and moving towards each other with the velocities $v_\mathrm{in}$ in the laboratory frame of reference. (The moving kinks can be obtained from the static by the Lorentz transformations.) Thus the initial conditions for our simulations were taken from the following expression:
\begin{equation}
\phi(x,t)=\phi_\mathrm{k}\left(\frac{x+\xi-v_\mathrm{in}t}{\sqrt{1-v_\mathrm{in}^2}}\right)+\phi_\mathrm{\bar{k}}\left(\frac{x-\xi+v_\mathrm{in}t}{\sqrt{1-v_\mathrm{in}^2}}\right)-2\pi.
\end{equation}

\begin{figure}[h!]
\subfigure[formation of a bion, $v_\mathrm{in}=0.2100$]{
\begin{minipage}{0.45\linewidth}
\centering\includegraphics[width=\linewidth]{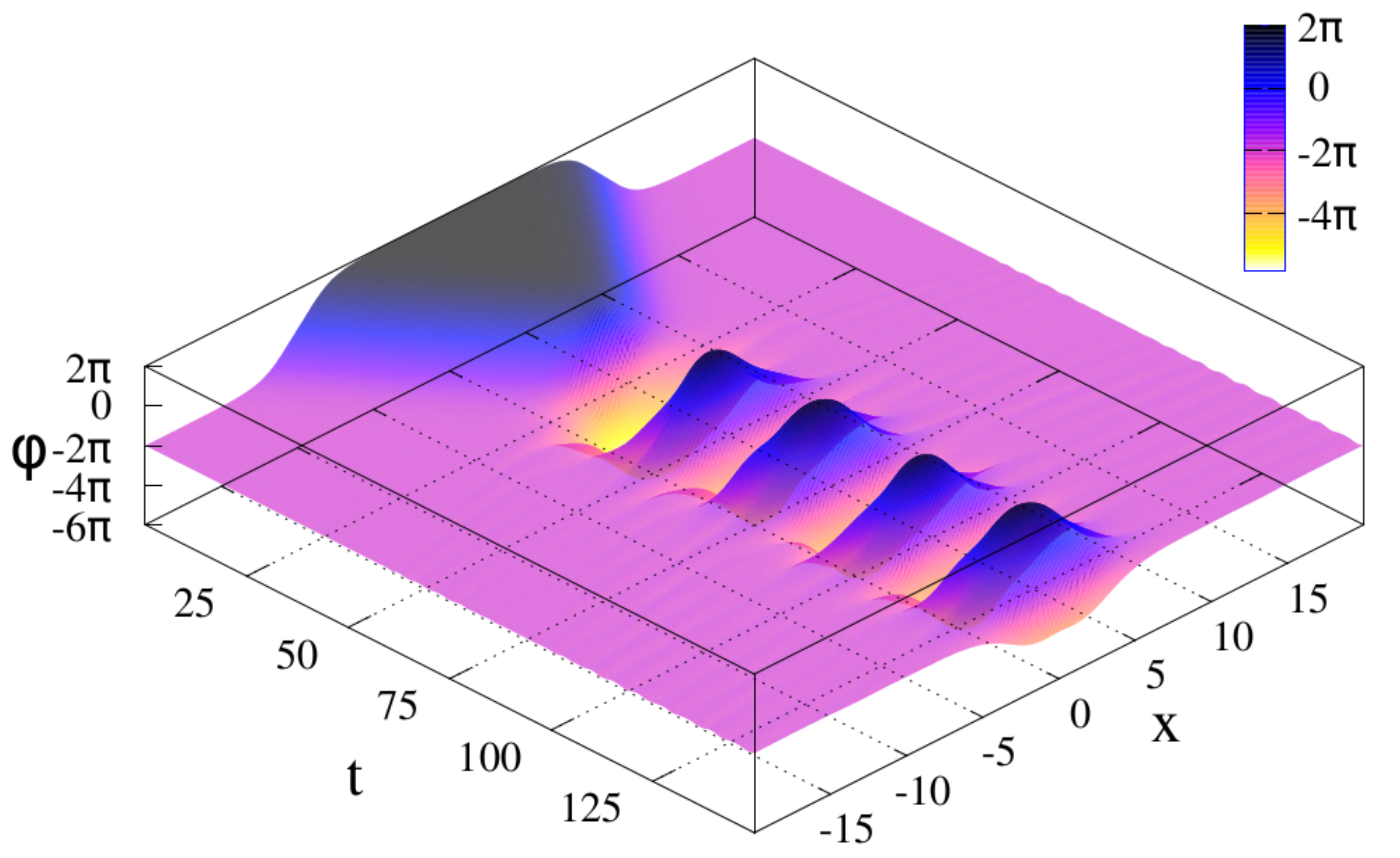}
\end{minipage}\label{fig:bion}
}
\hfill
\subfigure[escape of kinks, $v_\mathrm{in}=0.2500$]{
\begin{minipage}{0.45\linewidth}
\centering\includegraphics[width=\linewidth]{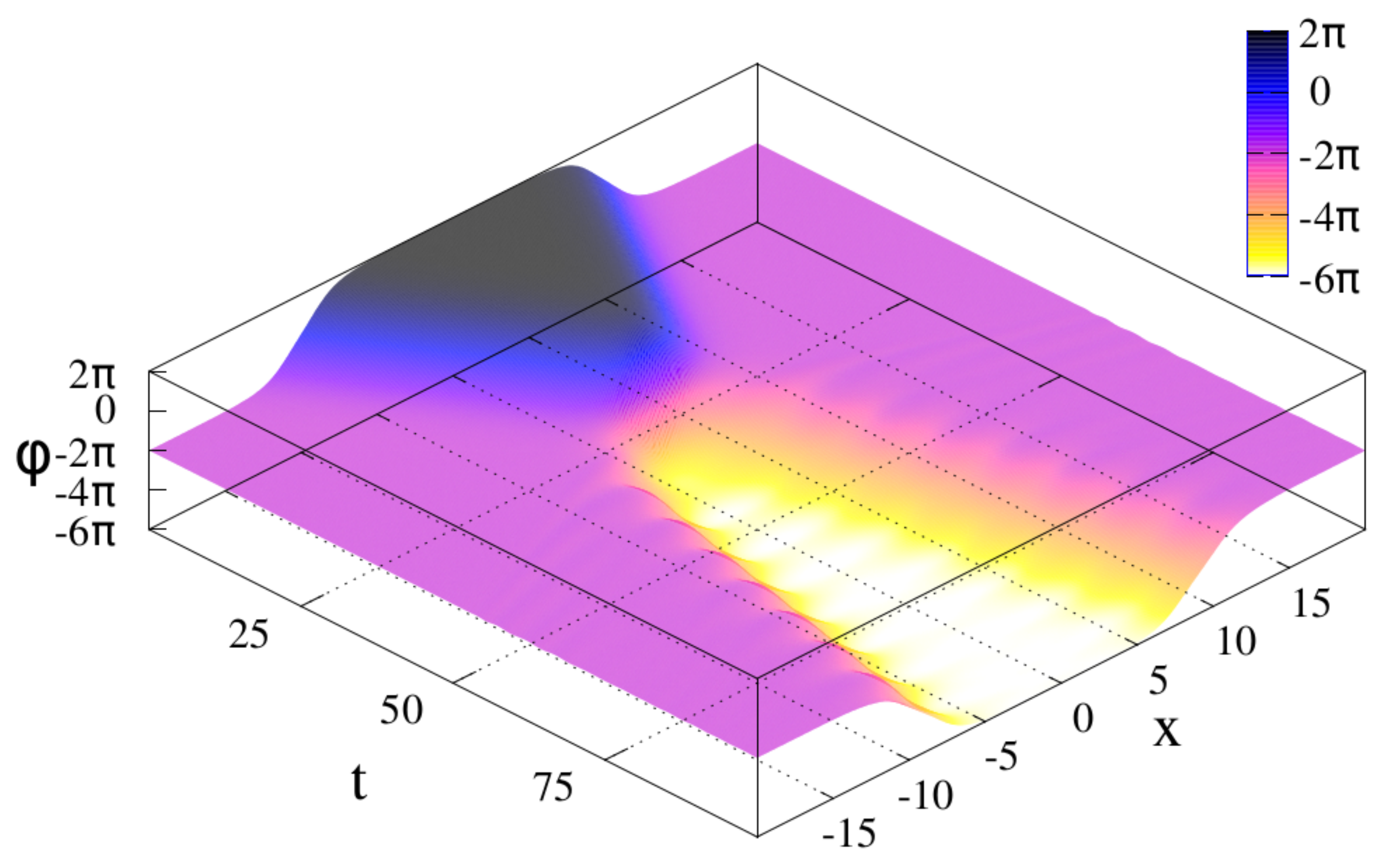}
\end{minipage}\label{fig:escape}
}
\caption{Space-time picture of the scattering for $R=1.0$.}
\end{figure}

In the numerical experiments we observed two different regimes of the collision: 1) at $v_\mathrm{in}<v_\mathrm{cr}$ the kink and antikink form {\it a bion} --- a bound state, which then decays slowly, emitting small waves, see figure \ref{fig:bion}; 2) at $v_\mathrm{in}>v_\mathrm{cr}$ the kink and antikink pass through each other and escape to spatial infinities, see figure \ref{fig:escape}.
The critical velocity $v_\mathrm{cr}$ depends on the parameter $R$, as it will be discussed below.

\begin{figure}[h!]
\subfigure[two-bounce window, $v_\mathrm{in}=0.2340$]{
\begin{minipage}{0.45\linewidth}
\centering\includegraphics[width=\linewidth]{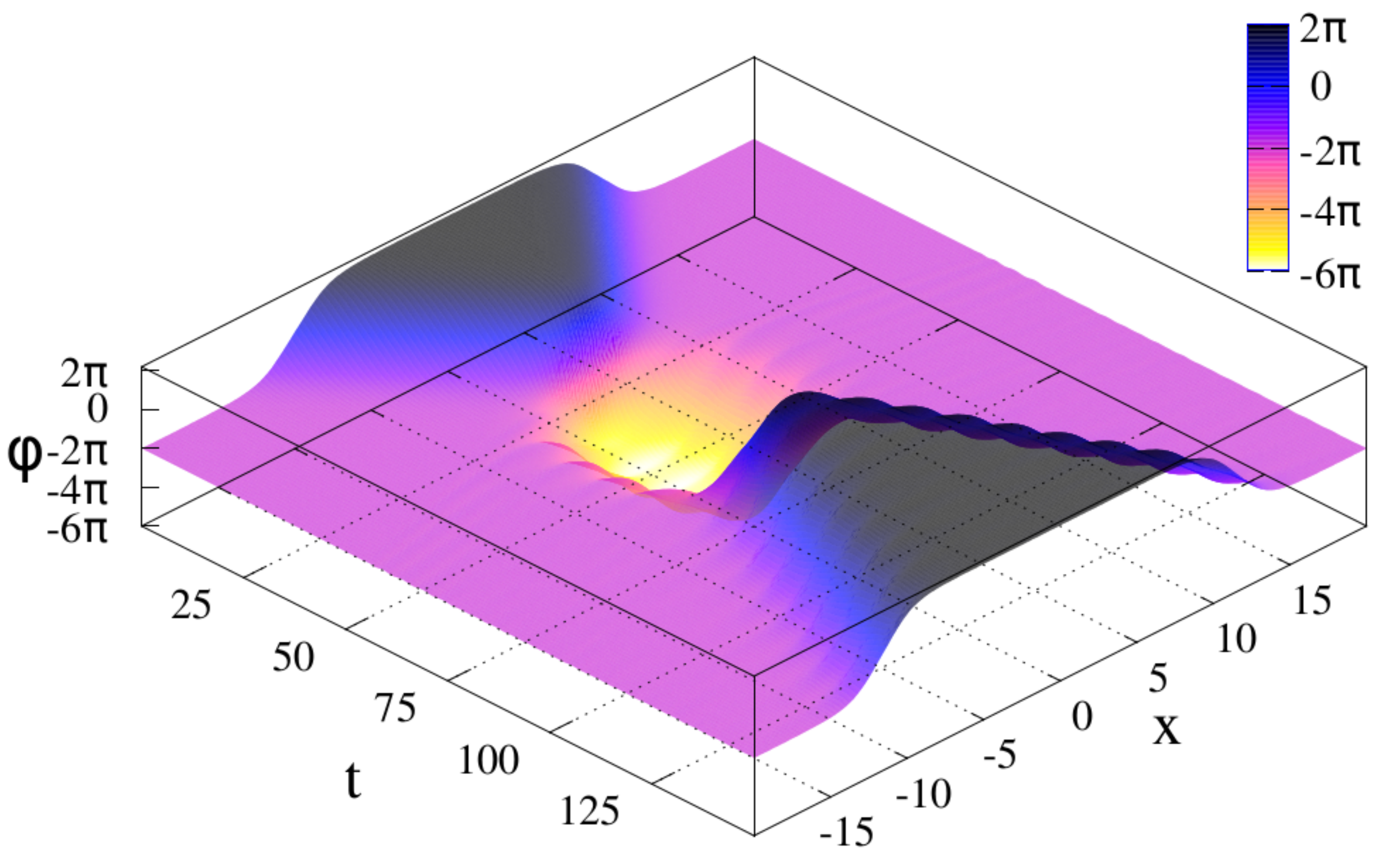}
\end{minipage}\label{fig:2win}
}
\hfill
\subfigure[three-bounce window, $v_\mathrm{in}=0.2380$]{
\begin{minipage}{0.45\linewidth}
\centering\includegraphics[width=\linewidth]{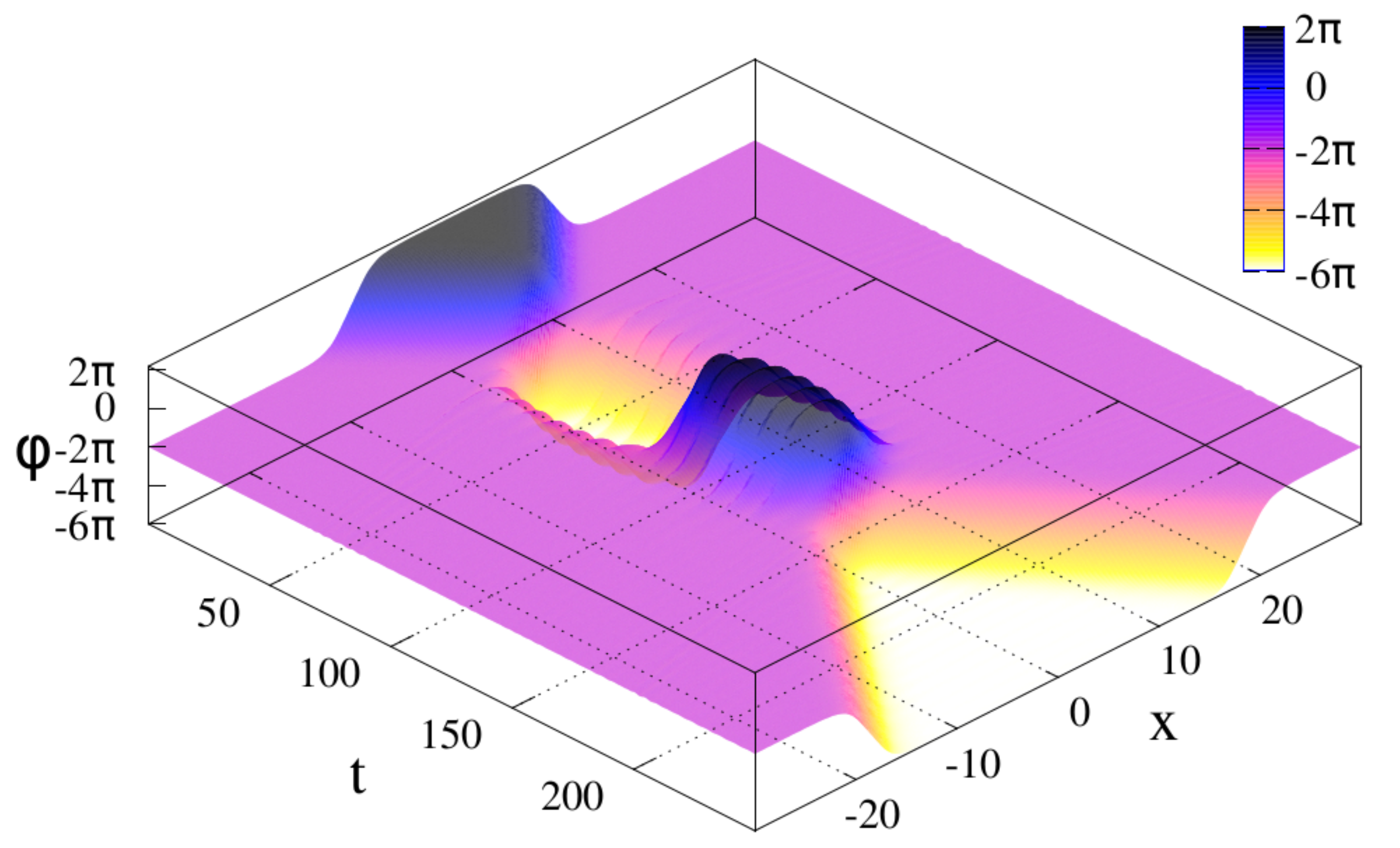}
\end{minipage}\label{fig:3win}
}
\caption{Escape of kinks after two and three collisions, $R=1.0$.}
\label{fig:windows}
\end{figure}

Besides that, at some initial velocities below the critical value we observed the so-called {\it escape windows} --- intervals of the initial velocity within which the kinks collide several times and then escape to infinities instead of forming a bound state, see figure \ref{fig:windows}.
This phenomenon can be explained by the resonant energy exchange between translational and vibrational modes of the kink (antikink), see, e.g., the review \cite{Kudryavtsev.UFN.1997.eng,Kudryavtsev.UFN.1997.rus}.

\begin{figure}[h!]
\centering\includegraphics[width=0.5\linewidth]{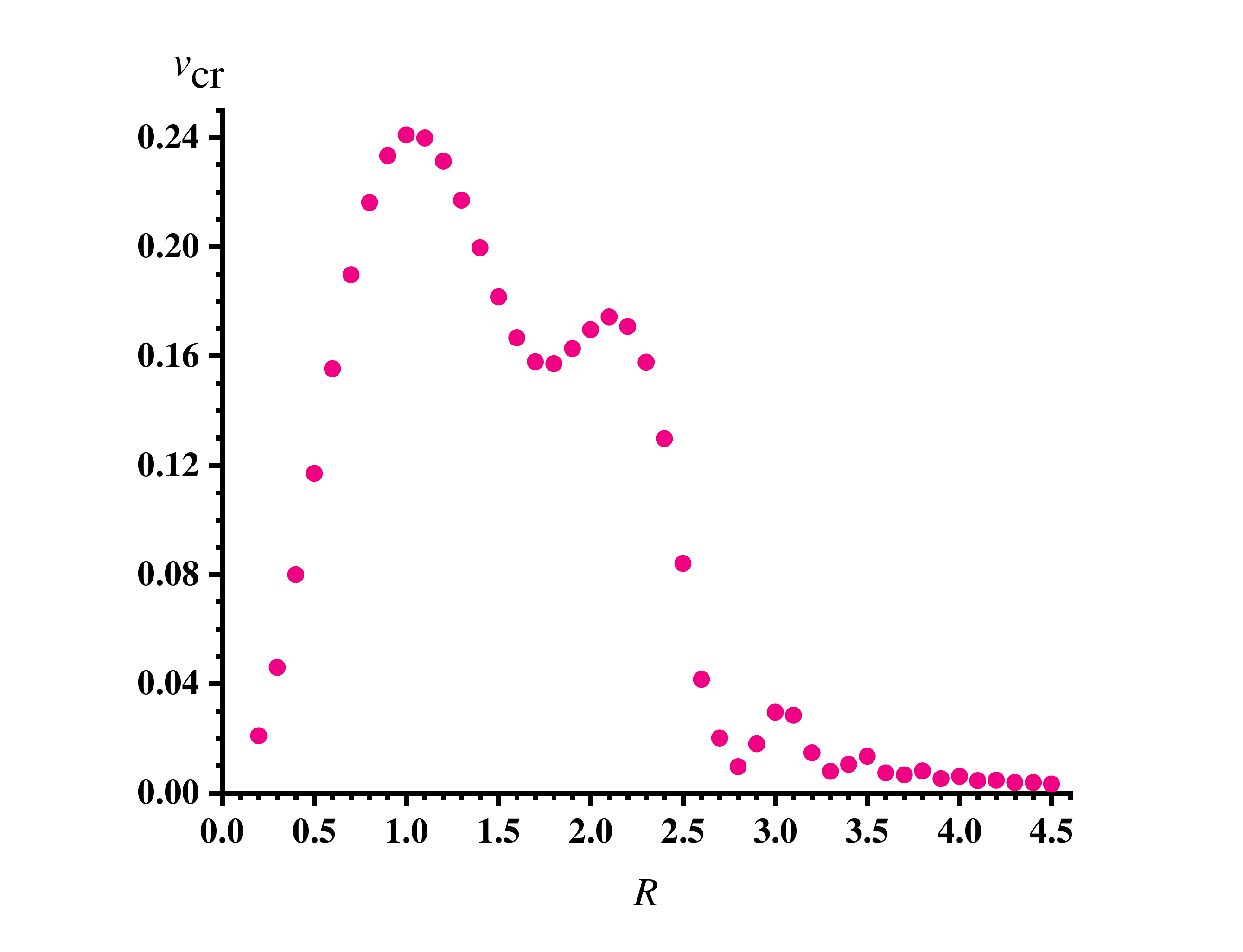}
\caption{$R$-dependence of the critical velocity.}
\label{fig:cr_R}
\end{figure}

We obtained the dependence of the critical velocity $v_\mathrm{cr}$ on the model parameter $R$, see figure~\ref{fig:cr_R}.
The curve $v_\mathrm{cr}(R)$ has several local maxima. (Note that this result has not been reported previously by other authors.) Such behavior could be a consequence of the complex structure of the DSG kink --- the colliding kink and antikink can be viewed as composed objects constructed from subkinks. Pairwise interaction of subkinks can lead to non-monotonous dependence of $v_\mathrm{cr}$ on the model parameter, see, e.g., \cite{Demirkaya.cc.2017}.

\begin{figure}[h!]
\subfigure[$v_\mathrm{in}=0.18472$]{
\begin{minipage}{0.45\linewidth}
\centering\includegraphics[width=\linewidth]{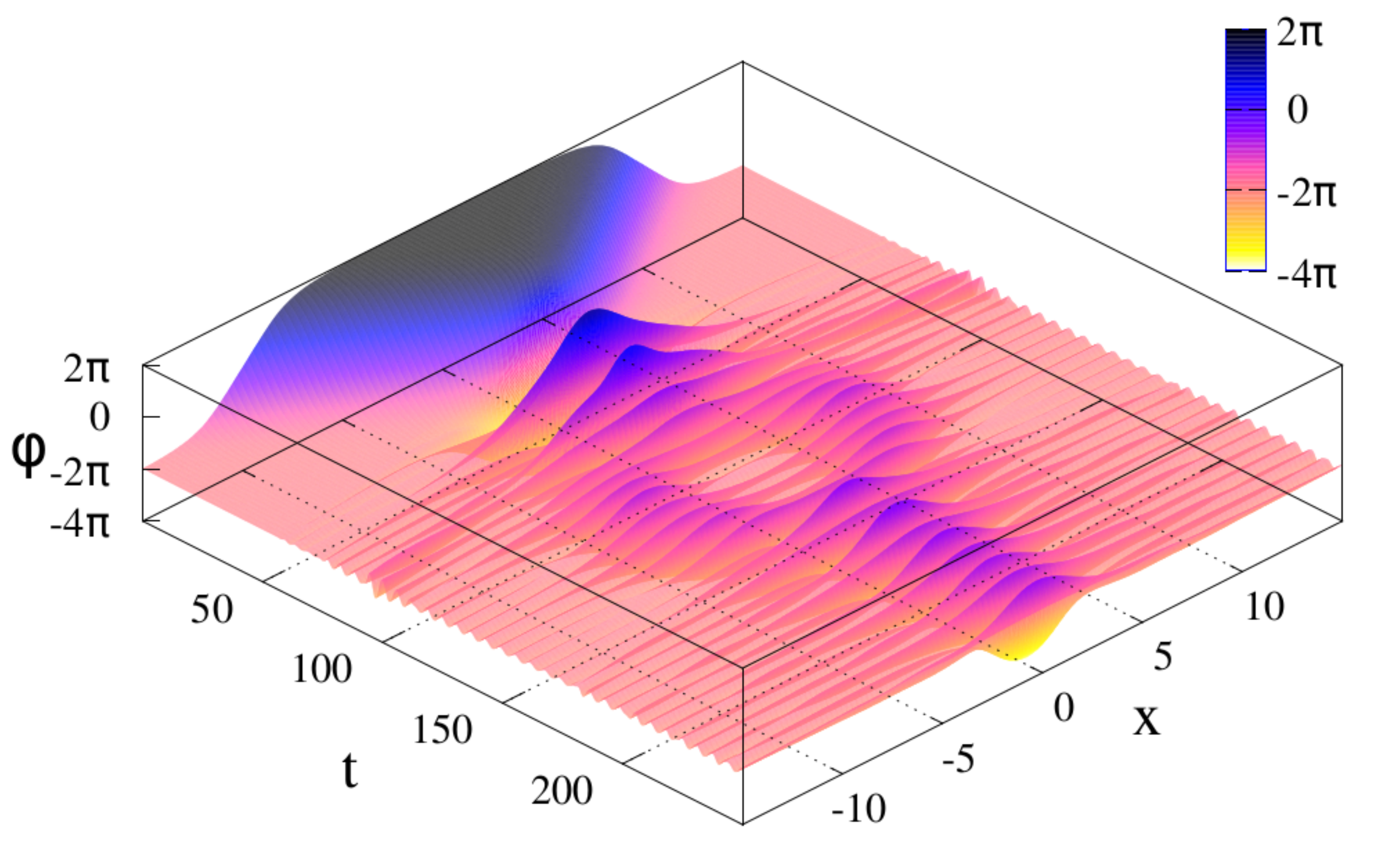}
\end{minipage}\label{fig:bion_osc}
}
\subfigure[$v_\mathrm{in}=0.18470$]{
\begin{minipage}{0.45\linewidth}
\centering\includegraphics[width=\linewidth]{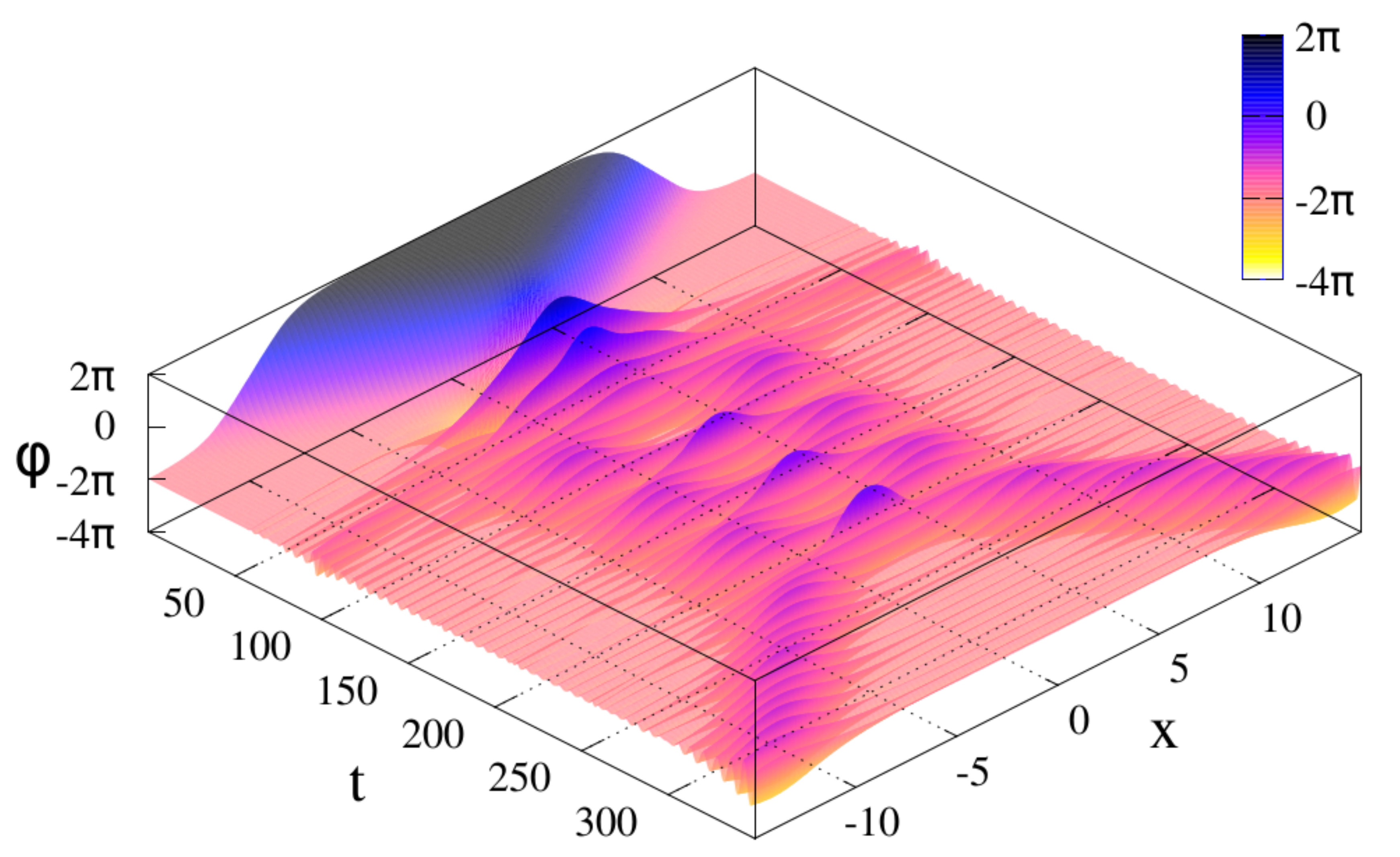}
\end{minipage}\label{fig:4win_osc}
}

\caption{Oscillons in the final state.}\label{fig:osc}
\end{figure}

Another interesting and new phenomenon that we have obtained is formation and further interaction of {\it oscillons}, see figure \ref{fig:osc}. These oscillating structures can form bound states (figure \ref{fig:bion_osc}), or even escape to spatial infinities after several collisions (figure \ref{fig:4win_osc}).

\section{Conclusion}

We have studied the scattering of the kink and antikink of the double sine-Gordon model. Our main new results are the following.
\begin{itemize}
\item
We have found that the critical value $v_\mathrm{cr}$ of the initial velocity depends on the model parameter $R$ non-monotonously. This fact has not been reported for the DSG model by other authors. We can assume that such behavior of $v_\mathrm{cr}(R)$ could be a consequence of pairwise interaction of subkinks.
\item
At the initial velocities below the critical value, $v_\mathrm{in}<v_\mathrm{cr}$, we observed new phenomenon --- formation of two oscillons in the final state. Their behavior seems to be rather complicated, they can form a bound state or, in some cases, can escape to spatial infinities after several collisions.
\end{itemize}

This research opens wide prospects for future work. For example, it would be interesting to study multikink collisions within the DSG model and to find maximal energy densities in such processes.

\section*{Acknowledgments}

The research was supported by the MEPhI Academic Excellence Project (contract No.\ 02.a03.21.0005, 27.08.2013).

\section*{References}

\bibliography{my_biblio}

\end{document}